# The missing pieces of the PuO$_2$ nanoparticle puzzle†

Evgeny Gerber, [a,b,c] Anna Yu. Romanchuk, [c] Ivan Pidchenko, [a,b]
Lucia Amidani, [a,b] Andre Rossberg, [a,b] Christoph Hennig, [a,b]
Gavin B. M. Vaughan, [d] Alexander Trigub, [e] Tolganay Egorova, [c]
Stephen Bauters, [a,b] Tatiana Plakhova, [c] Myrtille O. J. Y. Hunault, [f]
Stephan Weiss, [b] Sergei M. Butorin, [g] Andreas C. Scheinost, [a,b]
Stepan N. Kalmykov [c,e] and Kristina O. Kvashnina *[a,b,c]

The nanoscience field often produces results more mystifying than any other discipline. It has been argued that changes in the plutonium dioxide (PuO$_2$) particle size from bulk to nano can have a drastic effect on PuO$_2$ properties. Here we report a full characterization of PuO$_2$ nanoparticles (NPs) at the atomic level and probe their local and electronic structures by a variety of methods available at the synchrotron, including extended X-ray absorption fine structure (EXAFS) at the Pu L$_3$ edge, X-ray absorption near edge structure (XANES) in high energy resolution fluorescence detection (HERFD) mode at the Pu L$_3$ and M$_4$ edges, high energy X-ray scattering (HEXS) and X-ray diffraction (XRD). The particles were synthesized from precursors with different oxidation states of plutonium (III, IV, and V) under various environmentally and waste storage relevant conditions (pH 8 and pH > 10). Our experimental results analyzed with state-of-the-art theoretical approaches demonstrate that well dispersed, crystalline NPs with a size of ∼2.5 nm in diameter are always formed in spite of diverse chemical conditions. Identical crystal structures and the presence of only the Pu(IV) oxidation state in all NPs, reported here for the first time, indicate that the structure of PuO$_2$ NPs is very similar to that of the bulk PuO$_2$. All methods give complementary information and show that investigated fundamental properties of PuO$_2$ NPs, rather than being exotic, are very similar to those of the bulk PuO$_2$.

## Introduction

Plutonium (Pu) is one of the most complex and fascinating chemical elements in the periodic table.[1] At the end of 2014, there were about 2400 tonnes of irradiated and unirradiated plutonium from civilian nuclear power reactors located in 33 countries. Over the last ten years, this plutonium stock has grown at an average rate of almost 50 tonnes per year.[2] Nuclear reactors unavoidably generate nuclear waste, among which Pu isotopes are one of the longest-living. Most countries have adopted the concept of deep geological disposal of spent nuclear fuel and nuclear waste. Nuclear safety is of paramount importance and the main question, which remains unclear – is what will happen to Pu species present in the environment or nuclear wastes over the years? The path to answering these questions is by expanding the fundamental understanding of Pu properties with a special focus on the forms it can take in environmentally relevant conditions.

Pu has at least six oxidation states – from Pu(II)[3] to Pu(VII) – under certain conditions. In environmentally relevant conditions Pu may exist and even co-exist in four of them, from Pu(III) to Pu(VI).[4,5] Pu(IV) strongly hydrolyses in aqueous solutions resulting in the formation of polynuclear species, intrinsic colloids that may form a stable suspension or precipitates.[6,7] The speciation of Pu in aqueous solutions due to the complexity of its chemical behaviour (solubility and redox reactions[8,9]) combined with a great variety of environmental substrates (ligands, mineral interfaces, bacteria, solids, *etc.*) makes Pu geochemistry very complicated. It has been discovered that Pu may be transported by groundwater from contaminated sites on a scale of kilometres bounded with mineral[10,11] or organic colloids.[12] Recently, it was repeatedly found that $PuO_{2+x}$ NPs are formed during interfacial processes between Pu in different initial oxidation states and various mineral surfaces (hematite, goethite, quartz, and mica)[13–17] and bacteria.[18,19] All these results indicate the high importance of $PuO_2$ NPs in the context of environmental behaviour. Clearly, the variety of unknown or imprecisely known Pu species and crystal structures and their possible transformations make predictive modelling and extrapolation of the Pu chemical evolution and migration in the environment impossible or at least very challenging.

[a]*The Rossendorf Beamline at ESRF – The European Synchrotron, CS40220, 38043 Grenoble Cedex 9, France. E-mail: kristina.kvashnina@esrf.fr*
[b]*Helmholtz-Zentrum Dresden-Rossendorf (HZDR), Institute of Resource Ecology, PO Box 510119, 01314 Dresden, Germany*
[c]*Lomonosov Moscow State University, Department of Chemistry, 119991 Moscow, Russia*
[d]*ESRF – The European Synchrotron, CS40220, 38043 Grenoble Cedex 9, France*
[e]*National Research Centre "Kurchatov Institute", 123182 Moscow, Russia*
[f]*Synchrotron SOLEIL, L'Orme des Merisiers, Saint Aubin BP 48, 91192 Gif-sur-Yvette, France*
[g]*Molecular and Condensed Matter Physics, Department of Physics and Astronomy, Uppsala University, P.O. Box 516, Uppsala, Sweden*

†Electronic supplementary information (ESI) available. See DOI: 10.1039/d0nr03767b

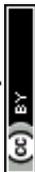

To improve our understanding of Pu properties, it is necessary to use licensed lab facilities to handle radioactive material safely combined with the most powerful experimental methods and complemented with theoretical advances. X-rays emitted by large-scale facilities – synchrotrons – can be used to study the atomic structure of such materials because of their penetrative behaviour and their sensitivity to the local and electronic structure of the selected element. Here we report a systematic investigation of $PuO_2$ NPs, synthesized using environmentally and waste storage relevant conditions, *i.e.* varying the pH (*e.g.* pH 8 and pH > 10) and the precursor (Pu(III), Pu(IV), Pu(V)). While pH 8 is more typical for far-field of nuclear waste disposal, an alkaline pH (>10) is possible in alkaline nuclear waste tanks and can also be reached in cement near-field environments.[20] The synthesized $PuO_2$ NPs are characterized by a variety of experimental methods: high-resolution transmission electron microscopy (HRTEM), selected-area electron diffraction (SAED), extended X-ray absorption fine structure (EXAFS), X-ray absorption near edge structure (XANES) in high energy resolution fluorescence detection (HERFD) mode, high energy X-ray scattering (HEXS), X-ray diffraction (XRD). The theoretical interpretation of the experimental data is based on the Anderson impurity model (AIM)[21] and *ab initio* calculations.[22] Fig. 1 shows a schematic illustration of the research methodology applied to $PuO_2$ NPs studies.

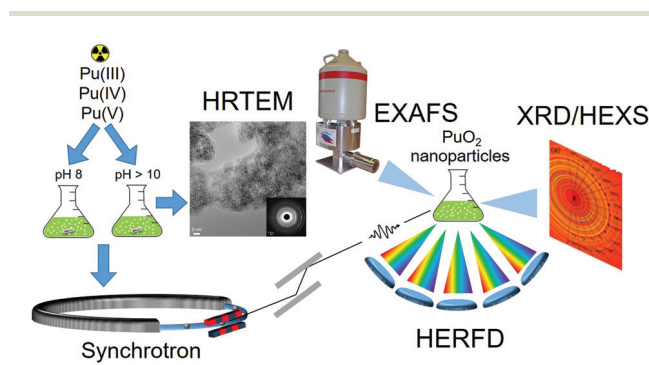

**Fig. 1** The schematic illustration of the research methodology applied to $PuO_2$ NPs studies. $PuO_2$ NPs synthesized from different precursors (Pu(III), Pu(IV) and Pu(V)) and at pH 8 and pH > 10 were analysed with HRTEM and various synchrotron techniques (EXAFS, XRD, HEXS and HERFD).

## Experimental section

### Synthetic procedures

Pu of 99.74 mass % of $^{242}$Pu was used for Pu stock. The Pu(III) solution was prepared from Pu stock solution by reduction with hydroxylamine hydrochloride when slightly heated in 1 M $HClO_4$. The Pu(IV) solution was obtained from Pu(III) oxidation with $NaNO_2$ in 5 M $HNO_3$. The Pu(V) solution was the result of the reaction of $H_2O_2$ with Pu(VI) solution (pH = 7), which in turn was also obtained from Pu stock by oxidation with $NaBrO_3$ under slight heating in 1 M $HClO_4$. All Pu valence states were verified by UV-vis spectrometry (TIDAS 100 J&M Analytics) and spectra of all initial solutions are listed in Fig. S1.† The presence of only Pu(VI) impurities in the initial solutions was measured to be less than 3%. The Pu(III), Pu(IV) and Pu(V) initial solutions (∼6 × 10$^{-4}$ M Pu) were divided into two parts to prepare $PuO_2$ NPs. The aliquots of Pu(III), Pu(IV) and Pu(V) from the first part were added to 3 M $NH_3·H_2O$ in the volume ratio 1 : 10 under continuous stirring. The pH of the 3 M ammonia solution is 12.5, but the pH was slightly decreasing during the synthesis procedure as the Pu cation hydrolyzed with acid formation. This set of samples would be called "from Pu(x) pH > 10", where x = III, IV, V. The aliquots of Pu(III), Pu(IV) and Pu(V) from the second part were added to water in the volume ratio 1 : 10 and then several drops of 3 M $NH_3·H_2O$ were added under continuous stirring to reach pH 8. The pH was also decreasing due to the Pu hydrolysis, hence additional drops of the ammonia solution were used to preserve pH 8. This set of samples would be called "from Pu(x) pH 8", where x = III, IV, V. Additional information about the synthesis is listed in ESI.† $PuO_2$ reference was purchased from Oakridge National Lab (Batch I.D. No. Pu-242-327A1).

### Methods

**HRTEM measurements.** The HRTEM images on a set of samples were recorded at Lomonosov Moscow State University (LMSU) with an aberration-corrected JEOL 2100F operated at 200 kV, yielding an information limit of 0.8 Å. The dark field (DF) images and energy-dispersive X-ray spectroscopy (EDX) analysis were performed in scanning transmission electron microscope (STEM) mode; the spot size was 1 nm with the HAADF and JED 2300 (JEOL) detectors.

**HEXS measurements.** High energy X-ray scattering data were collected at room temperature at the ID15A beamline of the European Synchrotron Radiation Facility (ESRF, Grenoble).[23] An incident energy of 120 000 eV was selected in order to be below the Pu K-edge at 121 791 eV and to minimize absorption. The K-edge XANES spectrum was measured to verify the incident energy (Fig. S2†). Data were collected up to 30 Å$^{-1}$ using a Dectris Pilatus 2 M CdTe pixel detector. Patterns were corrected for detector geometry, response and transparency, and integrated using a locally modified version of pyFAI[24] with outlier filtering. $F(q)$ and $G(r)$ were calculated from the resulting powder diffraction patterns using modules from DIFFPY-CMI[25] and locally developed cleaning algorithms. The full profile real-space refinement of crystal structures based on the pair distribution function $G(r)$ was made using PDFgui software.[26] PDFgui performs a least-squares refinement of the structural model to the $G(r)$. The parameters refined for the NPs series were lattice parameter $a$, particle diameter for the $G(r)$ shape damping function (spdiameter), the data scale factor, and factor which accounts for low-$r$ sharpening in $G(r)$ due to nearest-neighbour correlations (delta2). Parameters such as the $G(r)$ Gaussian dampening envelope due to limited $Q$-resolution and isotropic atomic displacement parameters (ADPs) were obtained from the fit of the experimental data of the $PuO_2$ reference and fixed at these values for the refinements of the NPs experimental sets, in order to minimize the number of refinable parameters and to obtain the most robust values for the coherent domain size. Free water was also included in the model in order to reproduce the contribution of water at the short-range order. All samples were fitted in the range from 1.7 to 20 Å and the maximum wave vector $Q$ of the data used for the generation of $G(r)$ was settled to 26 Å$^{-1}$. The $R_w$ value is a goodness of fit measure to show the agreement between calculated and experimental data.

**X-ray absorption near edge structure (XANES) in high energy resolution fluorescence detection (HERFD) mode at the Pu $L_3$ and $M_4$ edges and Pu $L_3$ extended X-ray absorption fine structure (EXAFS) spectroscopy.** The Pu $L_3$ HERFD experiments were performed at the Rossendorf Beamline at ESRF. The energy of the X-ray beam was tuned by a double-crystal monochromator operating in pseudo-channel-cut mode using a Si (111) crystal pair. Two rhodium-coated Si mirrors before and after the monochromator were used to collimate the beam and to reject higher harmonics. XANES spectra were simultaneously measured in total fluorescence yield (TFY) mode with a photodiode and in HERFD mode using an X-ray emission spectrometer.[27] The sample, each crystal analyser, and the silicon drift detector (Ketek) were positioned on their respective vertical Rowland circles of 0.5 m diameter. The Pu $L_3$ spectra were collected by recording the intensity of the Pu $L_{α1}$ emission line (~14 282 eV) as a function of the incident energy. The emission energy was selected using the [777] reflection of five spherically bent Si crystal analysers (with 0.5 m bending radius[28]) aligned at 75.7° Bragg angle. The intensity was normalised to the incident flux. A combined (incident convoluted with emitted) energy resolution of 2.8 eV was obtained as determined by measuring the full width at half maximum (FWHM) of the elastic peak.

The HERFD spectra at the Pu $M_4$ edge were collected at Beamline ID26 of ESRF.[29] The incident energy was selected using the [111] reflection from a double Si crystal monochromator. Rejection of higher harmonics was achieved by three Si mirrors at angles of 3.0, 3.5, and 4.0 mrad relative to the incident beam. The Pu HERFD spectra at the $M_4$ edge were obtained by recording the maximum intensity of the Pu $M_β$ emission line (~3534 eV) as a function of the incident energy. The emission energy was selected using the [220] reflection of five spherically bent Si crystal analysers (with 1 m bending radius) aligned at 66.0° Bragg angle. The paths of the incident and emitted X-rays through the air were minimized by employing a He filled bag to avoid losses in intensity due to absorption. A combined (incident convoluted with emitted) energy resolution of 0.4 eV was obtained.

One HERFD spectrum on $PuO_2$ NPs synthesized from Pu(III) precursor at pH 8 was collected at the MARS Beamline at SOLEIL using the [111] reflection from a double Si crystal monochromator. The Pu $M_β$ emission line was selected using one spherically bent Si crystal analyser (with 1 m bending radius) with [220] reflection. Experimental spectral broadening of HERFD data at the Pu $M_4$ edge collected at ID26 beamline was compared to one recorded at MARS beamline on $PuO_2$ reference sample and found to be identical (Fig. S3†).

The Pu $L_3$ edge EXAFS spectra were collected at the Rossendorf beamline (ROBL) in transmission mode using ion chambers. Spectra were collected at room temperature. Energy calibration was performed using the zero crossing of the second derivative of the K-edge of metallic Zr (~17 998 eV) and measured in parallel to the sample scans for each sample. Energy calibration and the averaging of the individual scans were performed with the software package SIXpack.[30] More information about EXAFS data analysis is in ESI.†

## Theory

The calculations of the Pu $M_4$ edge spectra were performed in the framework of the Anderson impurity model.[21] The spectra of bulk $PuO_2$ were calculated in a manner described in ref. 31 and 32 for the Pu(IV) system, taking into account the Pu 5f hybridization with the valence states and the full multiplet structure due to intra-atomic and crystal field interactions. The values of the Slater integrals obtained for Pu(IV) using the Hartree–Fock formalism were scaled down to 80% to account for the solid-state effect. Wybourne's crystal field parameters were set to $B_0^4 = -0.93$ eV and $B_0^6 = 0.35$ eV. The ground (final) state of the spectroscopic process was described by a linear combination of the $4f^4$ and $4f^5υ^1$ ($3d^94f^5$ and $3d^94f^6υ^1$) configurations where υ stands for an electronic hole in the valence level. The values for the model parameters were as following: the energy for the electron transfer from the valence band to the unoccupied Pu 5f level $Δ = 0.8$ eV; the 5f–5f Coulomb interaction $U_{ff} = 5.7$ eV; the 3d core hole potential acting on the 5f electron $U_{fc} = 6.5$ eV and the Pu 5f – valence state hybridization term $V = 1.10$ eV (0.9 eV) in the ground (final) state of the spectroscopic process.

The calculations of the $PuO_2$ $L_3$ edge spectra were performed using the FEFF9.6 code.[22] The input file was based on the atomic unit cell parameter $a = 0.540$ nm. Full multiple scattering (FMS) calculations were performed using the Hedin–Lundqvist exchange–correlation potential. The potential was calculated self consistently within a radius of 4 Å. The card UNFREEZEF was used to let the occupation of the f orbitals be calculated self-consistently. The core hole lifetime broadening was reduced by using the EXCHANGE card. A full multiple scattering (FMS) radius of 12 Å, selecting 75 atoms around the absorber was used. FEFF input file can be found in ESI, Fig. S4.†

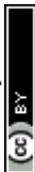

### Powder X-ray diffraction measurements

Powder X-ray diffraction (PXRD) data were collected at room temperature at the Rossendorf beamline of ESRF ($\lambda$ = 0.72756 Å, $2\theta_{max}$ = 59°). Samples were wet pastes enclosed in the three Kapton capillaries (triple confinement) with outer diameters of 0.81, 1.0, and 1.3 mm respectively. A diffractogram from an empty Kapton capillary was used in the background subtraction procedure for the samples. The lattice parameters were refined with WinCSD[33] by least-squares fitting. The FWHM and peak position were determined with TOPAS software.[34]

## Results and discussion

Pu compounds come in a variety of colours depending on the Pu oxidation state and the counter ion. We used acidic aqueous solutions of light blue Pu(III), yellow Pu(IV), and violet Pu(V) that were precipitated at pH 8 and pH > 10. Precipitates of Pu species of the same light green colour, typical to Pu(IV) solids, were formed from all solutions, suggesting that the electronic structure of the formed Pu materials might be very similar. However, it is unknown if the local structure could be disordered due to the possible presence of oxidized Pu(V) or reduced Pu(III) species. The particle size distribution and crystallinity could also differ depending on the synthetic route.

The HRTEM data reported in Fig. 2a confirm that similar NPs form (with respect to the size distribution and crystallinity), regardless of different precursors and pH conditions. A comparison of selected area electron diffraction (SAED, Fig. 2b) patterns with bulk $PuO_2$, as well as diffractograms from XRD measurements (*cf.* ESI†) show that the crystalline structure of the NPs is similar to that of bulk $PuO_2$. The particle size varies in the range of 2.3–3.2 nm, without apparent non-crystalline rim. Despite the small size of the NPs (extracted from XRD and reported in Fig. S5 and S6†), lattice parameters only differ slightly from bulk $PuO_2$ (Table 1), suggesting that Pu from the initial aqueous solutions transfers to $PuO_2$-like structure with Pu(IV) oxidation state.

Besides HRTEM and XRD there is another very powerful experimental method for investigating nano-scaled materials – HEXS analysis, as it can provide a fingerprint of the nanoparticle size and discriminate between short-range order (represented by finite non-random displacements from the ideal crystal structure) and random thermal displacements. The intensity profiles of the 2D diffraction for six investigated samples of $PuO_2$ NPs and a $PuO_2$ reference are shown in Fig. 2c. As discussed before (Fig. 2b), all diffraction patterns from $PuO_2$ NPs obtained by HEXS are similar, with the NPs having the same long-range structure as the $PuO_2$ reference with fcc structure, even though local structural deviations from the bulk should normally be observed in the nanosized materials with HEXS. The peaks are broader in the case of NPs, illustrating the effect of the nanosized coherent domains. The peak appearing at $Q \sim 2.40$ Å$^{-1}$ comes from the background (*i.e.* kapton capillary) and for most samples can be successfully subtracted using the measurement of the empty capillary. The corresponding reduced pair distribution function $G(r)$ is shown in Fig. 2d. Peaks from short-range correlations in $PuO_2$ NPs have the same sharpness as peaks of the $PuO_2$ reference, indicating that the short-range order is identical. The intensity of the oscillations drastically decreases with increasing $r$, nearly disappearing after $r$ = 20–25 Å; an indication of the coherence length of the $PuO_2$ particles. The damping of $G(r)$ can be refined in order to extract the average particle size, confirming the direct observation that the NPs diameters are in the range of 1.3–2.6 nm, in a good agreement with HRTEM and XRD estimations. The results of the full profile structural refinements are shown in Fig. S9 and

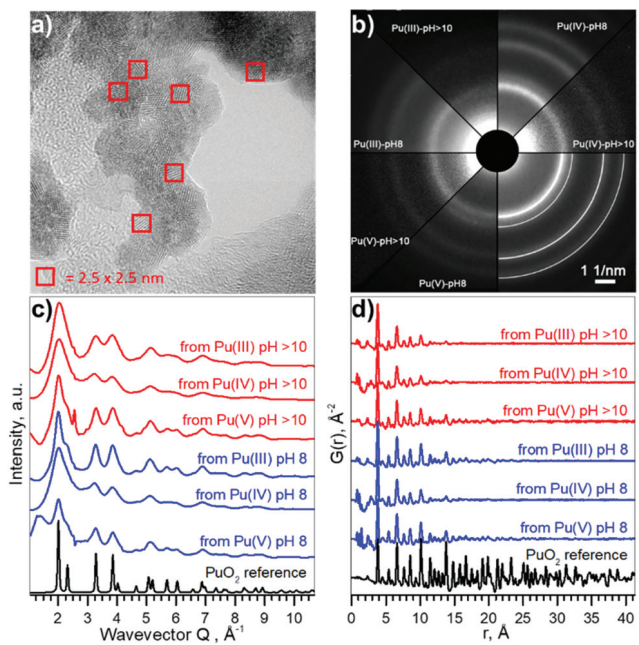

**Fig. 2** Size and structure information of $PuO_2$ NPs. (a) HRTEM image of NPs from Pu(V) pH 8, (b) SAED patterns of particles, white lines indicate peak positions for $PuO_2$ standard, (c) intensity profiles of the 2D diffraction of Pu samples, obtained by HEXS. (d) The corresponding reduced pair distribution functions $G(r)$ obtained by Fourier transformation (FT) of the data with $Q_{max}$ = 26.0 Å$^{-1}$. The magnified versions of a–b could be found in ESI as Fig. S7 and S8† respectively.

**Table 1** The comparison of the parameters particle sizes of NPs with different methods

| Sample | Diameter, nm | | | $a$, Å | |
|---|---|---|---|---|---|
| | HRTEM | XRD | HEXS | XRD | HEXS |
| From Pu(III) pH > 10 | 2.3(4) | 2.3(2) | 1.33 (5) | — | 5.384(6) |
| From Pu(IV) pH > 10 | 2.4(4) | 1.6(2) | 1.3(2) | 5.399(3) | 5.38(2) |
| From Pu(V) pH > 10 | 3.0(5) | 2.1(2) | 1.75(8) | 5.398(5) | 5.388(5) |
| From Pu(III) pH 8 | 2.7(4) | 2.2(3) | 2.07(8) | 5.404(4) | 5.402(4) |
| From Pu(IV) pH 8 | 3.2(8) | — | 1.5(2) | — | 5.39(2) |
| From Pu(V) pH 8 | 2.9(5) | 2.4(2) | 1.7(2) | 5.407(6) | 5.39(1) |
| $PuO_2$ reference | — | — | — | 5.4014(3) | 5.4032(3) |

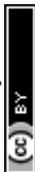

Table S1† for all $PuO_2$ NPs samples and for the $PuO_2$ reference, where bulk $PuO_2$ (space group $Fm\bar{3}m$) was used as the structure model.

Table 1 compares the results of the structural investigations of $PuO_2$ NPs by different methods: HRTEM, XRD, and HEXS. Although there are many factors that may influence the particle size estimation, the different independent methods, which we have implemented are all in a reasonable agreement. Table 2 summarizes the experimental methods and gives an overview of the information each method can supply. HRTEM is a local method, which gives information about particle size while XRD and HEXS provide information about coherent scattering domain size that sometimes may differ from particle size. The sizes calculated from HRTEM are slightly larger compared to other techniques, which might suggest that either there is a disordered or amorphous layer in $PuO_2$ NPs, restructured due to HRTEM deep vacuum conditions, or that the observed grains contain multiple overlapping subgrains. Low statistic of HRTEM methods may also influence on the result of size determination. Another possible explanation is that particles consisting of several crystallites may be counted as individual particles, resulting in underestimation of size in XRD and HEXS as they collect information about crystallite size.[35,36] HEXS allows for the characterization of short-range order and random thermal displacements. This method tends to find smaller particle sizes than those from XRD, although all fall within statistical uncertainty. Due to the agreement between several methods, we concluded that coherent diffracting crystalline domains and particle size are very close in our case (within the difference between XRD/HEXS and HRTEM values).

The most intriguing question concerning $PuO_2$ nanoparticles is the potential presence of various oxidation states of plutonium (referred to as $PuO_{2+x}$). Many authors[37,38] investigated the formation of $PuO_2$ NPs by various methods, but none of them gave an ambitious answer regarding the Pu oxidation state. In this work, we applied a new methodology and obtained straightforward information about Pu oxidation state by the HERFD method at the Pu $M_4$ edge.[39]

Pu complexity and variety of colours is determined by its ground state electronic configuration [Xe] $5f^6 7s^2$. We analysed $PuO_2$ NPs by the HERFD method at the Pu $L_3$ and $M_4$ edges, which is an element-selective technique as the energy of an absorption edge corresponds to the characteristic core-level energy. The information about the electronic structure, oxidation state and the local geometry of the absorbing atom can be obtained from the HERFD spectral shape – thanks to the possibility to record electronic transitions with high energy resolution by an X-ray emission spectrometer.[27] The short overview of the HERFD method is provided in Table 2.

Fig. 3 shows HERFD data recorded for six $PuO_2$ NPs samples at the Pu $L_3$ edge compared to both the spectrum of the $PuO_2$ reference and to the results of calculations. Electrons at the Pu $L_3$ absorption edge are excited from the core 2p level to the empty Pu 6d electronic levels.[39] All spectral features of $PuO_2$ NPs are very similar, corresponding to those of the $PuO_2$ reference and are well reproduced by calculations based on the bulk $PuO_2$ crystal structure (cf. Methods). There is a slight difference between all the samples and the reference at an incident energy of ~18 090 eV (first post-edge feature). The reason is that the first post-edge feature is very sensitive to the lowering of the number of coordinating atoms. A similar effect was found and explained for $ThO_2$ nanoparticles previously.[40] The shape of the spectra in the near-edge region is determined by Pu 6d electronic density of states, plotted at the bottom of Fig. 3. The Pu $L_3$ HERFD experiment was repeated after

Table 2  Short overview of the implemented analytical methods

| Analytical method | Information obtained | Comments |
|---|---|---|
| HRTEM | Information about morphology, particle size, elemental composition, and crystal structure with nano-resolution | Local method, which gives statistics only for a relatively small amount of particles; sample state can deviate during measurements due to the high vacuum and possible damage by the electron beam |
| XRD | Information about crystal structure and coherent scattering domains; statistically reliable; the ability to analyze mixed phases | Fast measurements (from seconds to minutes per sample); cannot easily distinguish between elements with similar $Z$; atomic level structure of amorphous substances can not be determined; nanoparticle peak broadening complicates the analysis |
| HEXS | Information about the local structure and coherent size domains; statistically reliable | Fast measurements (from seconds to minutes per sample); possibility to discriminate between short-range disorder and random thermal displacements; to distinguish amorphous, structurally disordered, and nanocrystalline materials |
| HERFD–XANES | Element selective method, which gives information about local site symmetry, oxidation state, and orbital occupancy | Fast measurements (from seconds to minutes per sample); any size of objects can be studied (from nm to cm); allows determination of low An concentrations; provides fingerprint information; quantitative information can be obtained |
| EXAFS | Information about local structure (bond distances, number and type of neighbours), sensitive to the type of absorbing element, local disorder and size effects | Time consuming measurements (from minutes to hours per sample); time consuming post-data collection analysis; provides only semi-quantitative estimations of oxidation state impurities; the results are severely dependent on the data quality and may be misinterpreted |

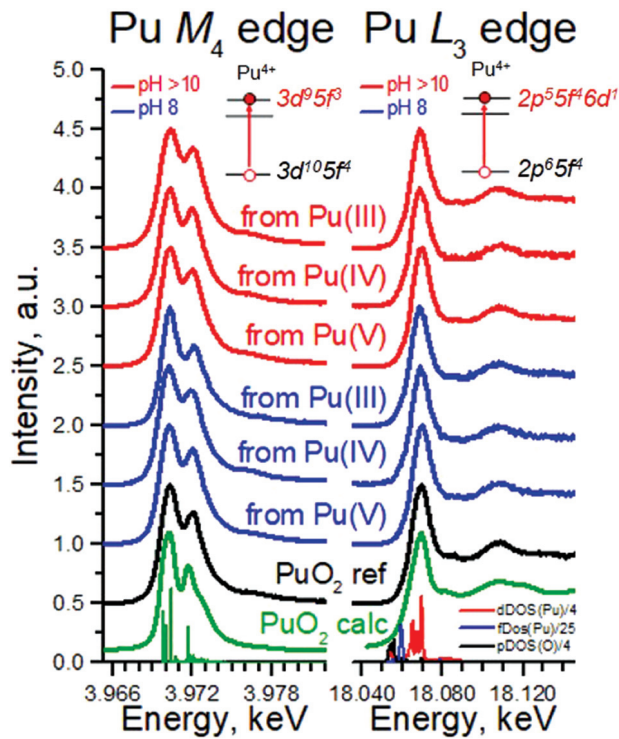

**Fig. 3** Pu $M_4$ and $L_3$ HERFD experimental data. Data recorded for six PuO$_2$ NPs samples and compared with a PuO$_2$ reference and theoretical calculations, using AIM approximation (left) and FEFF code (right). The density of states is plotted at the bottom for clarity.

4 months and no changes in the spectral shape were observed (Fig. S10†), confirming the stability of PuO$_2$ NPs over time.

To summarize, HERFD results at the Pu $L_3$ edge confirm that the local environment of all PuO$_2$ NPs samples is similar to bulk PuO$_2$, in agreement with the previous HRTEM, XRD, and HEXS results. This is also supported by a comparison of the area under the white line (WL) and its FWHM (Table S2†). Moreover, the HERFD data confirm the presence of the Pu(IV) oxidation state as the dominant valence of PuO$_2$ NPs, as seen from the positions of the Pu white line. A minor WL energy shift among the samples is observed (within 0.2–0.4 eV), which prevents exclusion of tiny amounts of other oxidation states being present. A close inspection of the density of states, shown in the bottom of Fig. 3 (right), reveals that the Pu 5f density of states is much closer to the Fermi level than that of the Pu 6d states; therefore probing the Pu 5f density of states is more suited to investigate the electronic structure of actinides (An). Previous experiments showed that HERFD measurements at the An $M_4$ edge, characterized by the An 3d–5f electronic transitions, are very powerful for oxidation state identification.[39,41]

Fig. 3 (left) shows the Pu $M_4$ edge measurements on six PuO$_2$ NPs samples compared to both the spectrum of a PuO$_2$ reference and to the results of calculations based on Anderson impurity model (AIM). AIM fully accounts for electron correlations and treats the inter-atomic interactions as a pertur-bation. This approach was already shown to be the best for analysis of the HERFD data at the An $M_4$ edge.[31,39,41] The shape and the position of the main WLs in the Pu $M_4$ edge spectra of PuO$_2$ NPs are similar to those in the spectrum of the PuO$_2$ reference. The HERFD spectrum of PuO$_2$ shows a sharp peak at ~3970 eV due to the transitions from the 3d$_{3/2}$ core level to the unoccupied 5f$_{5/2}$ level and a shoulder at higher energy. This splitting of the Pu $M_4$ main edge transitions is reproduced very well by theoretical calculations and relates to the multiplet splitting of the Pu 5f states.[39] These results confirm the presence of only the Pu(IV) oxidation state in PuO$_2$ NPs and settle the controversial debates in the last years about the presence of other oxidation states. There are reports about the presence of Pu(III),[42] Pu(V),[9,43] and Pu(VI)[44] in PuO$_{2\pm x}$ species. Contrary to that, recent studies of Pu nanoclusters claim the presence of only the Pu(IV)[45] oxidation state. These results were obtained on Pu dioxide species synthesized under different conditions (pH range, concentrations *etc.*). We believe that all previously obtained results might be verified by the newly available HERFD method at the Pu $M_4$ edge.

In general, spectroscopic methods with standard resolution have been heavily used in studies of An systems in the past years. One of the most popular methods is EXAFS at the An $L_3$ edge, (information about this method is in Table 2). Pu $L_3$ edge EXAFS investigations of PuO$_2$ NPs have been carried out multiple times by different research groups,[38,43,46–50] but results are heavily debated and are not in agreement with one another. The main discrepancy is related to the different interpretations of the first coordination sphere, which refers to Pu–O interaction and can be classified as:

(1) a single Pu–O interaction similar to the one in PuO$_2$;

(2) several Pu–O interactions (with different coordination of O atoms);

(3) several Pu–O interactions, including Pu in various oxidation states.

For example, Conradson *et al.*[43] claimed that PuO$_2$ rarely exists and proposed as a description of the PuO$_2$ phase PuO$_{2+x-y}$(OH)$_{2y}$·zH$_2$O with a variety of Pu–O distances and most likely containing Pu(V). It was reported that EXAFS data shows at least three different oxygen scattering paths between 1.8 and 2.4 Å, however, the distances of all three paths vary significantly between 1.85–1.96, 2.13–2.28, and 2.26–2.39 Å for excess O (PuO$_2^+$ moieties), Pu–OH and Pu–O respectively.

Rothe *et al.*[46] found only Pu(IV) during PuO$_2$ intrinsic colloid formation and explained the observed splitting of the Pu–O coordination sphere by condensation of Pu$_n$O$_p$(OH)$_{4n-2p}$(H$_2$O)$_z$ units, with the shorter 2.20–2.24 Å Pu–O distances corresponding to surface Pu–OH groups and the longer 2.38–2.42 Å distances to Pu–O groups. Bonato *et al.*[38] recently reported the splitting of the first oxygen shell in PuO$_2$ nanoparticles and explained it with a disorder of the local structure at the surface of the NPs. Authors fixed distance parameters and coordination numbers for the first coordination shell, and refined only the Debye–Waller factor. In order to confirm such a splitting of the Pu–O shell previously observed by other authors,[38,43,46] we performed here several analyses of

 

EXAFS data, using one, two, and three different Pu–O shells, with a refinement of Debye–Waller (DW) factors, distances and coordination numbers, assuming that nanosize effects and disorder may impact simultaneously the resulting values.

The one oxygen-shell fit is based on the assumption of a $PuO_2$-like structure of the samples and includes one Pu–O and one Pu–Pu interaction. Fig. 4a shows Pu $L_3$-EXAFS shell fit results, which indicate this approach can be implemented for all samples assuming a $PuO_2$-like structure, with characteristic distances 2.31 Å and 3.81 Å for Pu–O and Pu–Pu respectively. The best fit parameters are listed in Table 3. In comparison to bulk $PuO_2$, the fit parameters of the NPs show significant differences. First, the DW for the Pu–O shell is higher (note that 0.01 Å$^2$ was imposed as an upper limit to prevent collinearity effects and hence unreasonable values during the least-square refinement[51]), indicative of a higher static disorder due to a larger fraction of surface oxygen atoms. Second, the NPs Pu–Pu CN is lower and their DW is higher than those of the $PuO_2$ reference, again indicative of a larger fraction of under-coordinated Pu atoms at the surface with the higher static disorder. These observed effects of increased static disorder and reduced coordination numbers are clear influences of the nanometer-size of the particles, without the need to invoke different Pu oxidation states or a substantially different structure. In line with the very similar particle sizes of all NPs samples, their shell fit values remain remarkably similar, while they are significantly different from those of the bulk sample.

Both two and three shell fit models were based on previously published fit attempts. Rothe et al.[46] observed oxygen shell splitting and used a two shell fit model with a dominant oxygen contribution at average distances slightly larger than that for bulk crystalline $PuO_2$ i.e. $PuO_2(cr)$ (2.38–2.43 Å) and a smaller contribution at a significantly shorter distance (around 2.22 Å) and assigned this shorter Pu–O distance to hydroxyl groups. Our three shell fit follows the results of Conradson et al.,[43] where also Pu(v)–O distances in the range of 1.85–1.96 Å were fitted. We applied these assignments to our models and kept the original species assignments for the sake of comparison with previous reports. The results of these approaches are described in ESI (Fig. S11–S14 and Tables S3, S4†).

Similar approaches were used previously for the characterization of Pu colloids.[38,48] We used here the combination of techniques (HERFD, EXAFS, and HEXS) and theoretical calculations in order to find the most consistent approach. The three shell fit approach can be discarded due to the contradiction between EXAFS and HERFD data. The first Pu–O path of 1.83 Å has been previously assigned to the $PuO_2^+$ moiety, with pentavalent Pu reaching up to 15%, which is not consistent with the HERFD results at the Pu $M_4$ edge (more info in ESI†). Since we cannot discard the presence of several Pu(iv)–O distances with HERFD method, we analyzed the EXAFS data in addition to the shell fitting approach with the Landweber iteration (LI),[52,53] which is used to derive the radial particle distribution function ($n(r)$) directly from EXAFS without the need to apply a shell fitting scheme based on the symmetric (Gaussian) distribution of interatomic distances. The advantage of this method is hence the possibility to reconstruct also the asymmetric $n(r)$ directly from the EXAFS spectra. In line with the EXAFS shell fit, this LI approach showed only one Pu–O contribution (Fig. S12c†). We also calculated a theoretical model EXAFS spectrum (Fig. S13a†) based on the EXAFS parameters given by Rothe et al.[46] to confirm the ability of the LI approach for resolving close Pu–O contributions as described in ESI.† In addition, we performed Monte–Carlo (MC) EXAFS simulations[54–60] as a complementary method to LI. The MC simulation resulted in a very good agreement with the experimental EXAFS spectra (Fig. S14a†) and again the symmetric shape of the first shell Pu–O $n(r)$ is observed (Fig. 4b and Fig. S14c†), thus only one Pu–O contribution is present in our system. In conclusion, the agreement among the results from the EXAFS shell fit, MC EXAFS simulations (Fig. 4 and Fig. S14†), EXAFS LI approach and from HEXS all unequivocally indicate that only one Pu(iv)–O interaction with a rather symmetric distance distribution (similar to the $PuO_2$ reference) is present for all six $PuO_2$ NPs samples.

These observations lead us to conclude that the best model is a $PuO_2$ structure with Pu–O and Pu–Pu scattering paths to describe the features of our samples. However, we cannot claim that there are no hydroxyl groups in the outer shell of NPs, because Pu in aqueous solutions is always hydrolyzed, which means a variety of different kinds of hydrolyzed forms may exist during the formation of NPs. Nevertheless, these conclusions from the oxygen shell may help elucidate the mechanisms of the $PuO_2$ NPs formation.

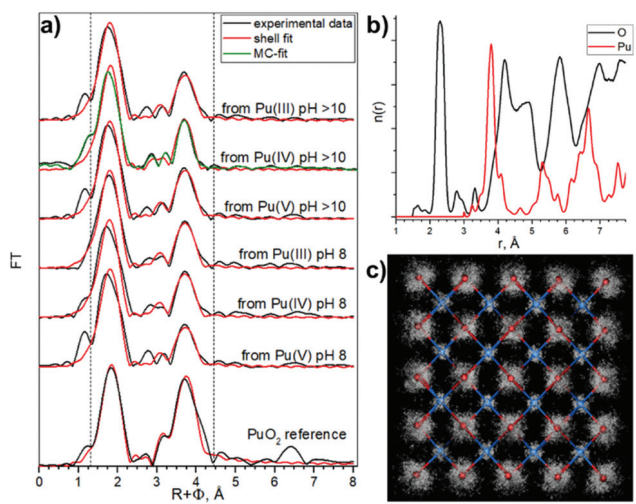

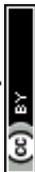

**Fig. 4** Pu $L_3$ EXAFS results. (a) Pu $L_3$ EXAFS spectra $\chi(R)$ fit results, Fourier transform (FT) magnitude of experimental EXAFS data (black) and a one Pu–O, one Pu–Pu shell fit (red), Monte–Carlo simulations fit (green). The area of the shell fit in $R$-space is marked with two dashed lines, (b) MC simulation output: radial particle distribution function ($n(r)$) for Pu–O and Pu–Pu, (c) 3D structural refinement based on bulk $PuO_2$ structure.

Table 3 Metric parameters extracted by least-squares fit analysis of Pu L$_3$ EXAFS spectra with one Pu–O and one Pu–Pu paths ($k$ range of 2.0–11.5 Å$^{-1}$)$^a$

| Sample | First coordination shell | | | Further shells | | | $\Delta E_0$ [eV] | $\chi^2_{res}$ % |
| --- | --- | --- | --- | --- | --- | --- | --- | --- |
| | CN | $R$ [Å] | $\sigma^2$ [Å$^2$] | CN | $R$ [Å] | $\sigma^2$ [Å$^2$] | | |
| From Pu(III) pH > 10 | 7.5 O | 2.31 | 0.0100 | 3.6 Pu | 3.80 | 0.0045 | 5.1 | 14.0 |
| From Pu(IV) pH > 10 | 7.5 O | 2.30 | 0.0100 | 4.3 Pu | 3.79 | 0.0071 | 4.9 | 14.8 |
| From Pu(V) pH > 10 | 7.8 O | 2.31 | 0.0100 | 3.8 Pu | 3.81 | 0.0041 | 5.1 | 12.3 |
| From Pu(III) pH 8 | 7.5 O | 2.31 | 0.0100 | 4.5 Pu | 3.81 | 0.0060 | 4.7 | 12.9 |
| From Pu(IV) pH 8 | 7.2 O | 2.30 | 0.0100 | 4.5 Pu | 3.80 | 0.0065 | 4.9 | 15.5 |
| From Pu(V) pH 8 | 7.7 O | 2.32 | 0.0100 | 3.2 Pu | 3.81 | 0.0043 | 5.2 | 16.5 |
| PuO$_2$ reference | 8.7 O | 2.33 | 0.0068 | 6.7 Pu | 3.82 Pu | 0.0011 | 7.3 | 9.0 |
| PuO$_2$ structure | 8 O | 2.337 | | 12 Pu | 3.817 Pu | | | |

$^a$ CN: coordination number with error ±25%, $R$: Radial distance with error ±0.01 Å, $\sigma^2$: Debye–Waller factor with error ±0.0005 Å$^2$.

There are two proposed mechanisms for PuO$_2$ NPs formation: the first one involves the condensation of [Pu(OH)$_n$]$^{(4-n)+}$ to yield hydroxo-bridged species through an olation reaction[7]

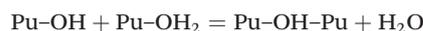

$$\text{Pu–OH} + \text{Pu–OH}_2 = \text{Pu–OH–Pu} + \text{H}_2\text{O}$$

According to this theory, oxo-hydroxo particles are formed as intermediate products. In this case several Pu–O contributions with slightly different Pu–O distances should be present[43,61] (such as native Pu–O as well as Pu–O in the terminal positions, Pu–OH and Pu–OH$_2$) in the EXAFS spectra, hence oxygen shell splitting should appear. The other proposed mechanism is that PuO$_2$ NPs form directly by an oxolation reaction[62,63]

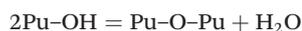

$$2\text{Pu–OH} = \text{Pu–O–Pu} + \text{H}_2\text{O}$$

hence no oxyhydroxide or hydrous oxides are expected in the inner core of NPs. A bond-length distribution can still result from lattice distortions and different Pu local environments in the inner core and on the surface.[63] However, Monte–Carlo simulations reproduce the distortion effect, where no bond-length distribution appears. Since there is neither evidence of oxygen shell splitting nor Pu–O distances other than the Pu–O oxide distance of the native cubic structure, it is hardly possible to accept that the olation mechanism is responsible for the polymerization reaction leading to the formation of PuO$_2$ NPs.

Condensation of Pu(IV) by olation reactions is not prevalent in the literature; there is only one published report characterizing a solid-state structure comprised of hydroxobridged oligomers.[64] It is known that Th(IV), which is the softest among the tetravalent ions in Pearson's acid–base concept, is present as a [Th$_2$(OH)$_2$]$^{6+}$ cluster in both aqueous solution and solid precipitates,[65,66] but other mononuclear and polynuclear species, with two, four and six atoms of Th in the cluster also exist.[67,68] The hardest tetravalent ions stable in aqueous solution, Zr(IV) and Hf(IV), have dihydroxo-bridged tetramers [M$_4$(OH)$_8$(H$_2$O)$_{16}$]$^{8+}$ (M = Zr, Hf) as the dominant solution species,[69] though many others exist under certain conditions[70] Similar well-defined Pu polymers have not been found.[71] Ce(IV) and Pu(IV) have a charge-to-radius ratio intermediate to and bounded by Th(IV) and Hf(IV) and consequently their hydrolysis chemistry is similar. For Ce and Pu, predominantly oxo-bridged species have been conclusively isolated.[63,72] However, after the discovery of hydroxo/oxo-bridged hexanuclear complexes for Pu(IV) and Ce(IV), [M$_6$(μ$_3$-O)$_4$(μ$_3$-OH)$_4$]$^{12+}$ (M = Pu, Ce)[73,74] and two dihydroxo-bridged Pu(IV) dimers,[75] it became clear that a simple hard and soft acid-base concept is not sufficient to predict the hydrolysis chemistry of these metal ions. The real mechanism of Pu hydrolysis appears to be much more complicated than simple olation or oxolation.[76]

## Conclusions

In this paper we report a full characterization of PuO$_2$ NPs at the atomic level and refine the crystal and electronic structures by advanced synchrotron-based methods. Six samples of PuO$_2$ NPs were synthesized under environmentally and waste storage relevant conditions – at pH 8 and pH > 10 from Pu(III), Pu(IV) and Pu(V) precursors. Despite varying synthesis conditions, the Pu oxidation state in the PuO$_2$ NPs turns out to be exclusively Pu(IV), as proven here by HERFD method at the Pu M$_4$ edge for the first time. The array of complementary methods applied in this work, namely EXAFS, HEXS, XRD, and HRTEM, support the conclusion that PuO$_2$ NPs are surprisingly homogeneous and monodisperse. It is surprising and not understood why always NPs of ~2.5 nm are formed. There are certain difficulties in obtaining PuO$_2$ NPs of bigger size (5–10 or 15 nm) via the chemical precipitation method. All collected information plays an important role in explaining plutonium chemistry under real conditions. It will help to advance, to model and to predict long-term Pu release from deep underground repositories of nuclear waste and contaminated sites.

## Author contributions

K. O. K and S. N. K planned and supervised the project. A. Yu. R., E. G., I. P. and S. W. performed synthesis of PuO$_2$ NPs. T. E. performed HRTEM characterizations of the samples. K. O. K.,

A. Yu. R., E. G., I. P, A. T. carried out XRD, EXAFS and HERFD measurements at the Pu $L_3$. C. H. and E. G. carried out XRD measurements and analysed data. K. O. K., E. G., I. P. and L. A. performed HERFD experiments at the Pu $M_4$ edge. M. O. J. Y. H., S. B. and T. P. performed Pu $M_4$ HERFD measurements on one $PuO_2$ NPs sample, synthesized from Pu(III) precursor at pH 8 at MARS beamline. A. C. S., A. R., A. T. and E. G. performed EXAFS experiment and analysed EXAFS data. G. V., and E. G. carried out and performed the analysis of HEXS data. S. M. B. did theoretical simulations of Pu $M_4$ data. K. O. K., E. G., A. Yu. R., G. V., A. C. S., S. M. B. and S. N. K. co-wrote the paper. All authors discussed the results and contributed to the final manuscript.

## Conflicts of interest

There are no conflicts to declare.

## Acknowledgements

This research was funded by European Commission Council under ERC [grant N759696]. Authors thank HZDR for the beamtime allocation at BM20 beamline and thank ESRF for beamtime allocation at ID15a and ID26 beamlines. We thank SOLEIL synchrotron for providing beamtime. We acknowledge help of P. Glatzel and T. Bohdan at ID26 beamline of ESRF during the HERFD experiment at the Pu $M_4$ edge. The authors would like to thank P. Colomp and R. Murray from the ESRF safety group for their help in handling radioactive samples at ID26, ID15a and ROBL beamlines. A. Yu. R. acknowledges support from RFBR according to the research project No. 16-33-60043 mol_a_dk. E. G. and S. N. K. acknowledge support by the Russian Ministry of Science and Education under grant no. 075-15-2019-1891. S. M. B. acknowledges support from the Swedish Research Council (research grant 2017-06465).

## Notes and references

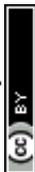